\documentclass[twocolumn,aps]{revtex4}
\usepackage[]{fontenc}
\usepackage[latin1]{inputenc}
\usepackage{amsmath}
\usepackage{graphicx}
\usepackage{amssymb}

\begin{document}

\title{In-Chain Tunneling Through Charge-Density Wave Nanoconstrictions
and Break-Junctions}

\author{K. O'Neill}

\affiliation{Kavli Institute of Nanoscience Delft, Delft University of Technology,
Lorentzweg 1, 2628 CJ Delft, The Netherlands}

\author{E. Slot}

\affiliation{Kavli Institute of Nanoscience Delft, Delft University of Technology,
Lorentzweg 1, 2628 CJ Delft, The Netherlands}

\author{R. E. Thorne}

\affiliation{Laboratory of Atomic and Solid State Physics, Cornell University,
Ithaca, New York 14853}

\author{H. S. J. van der Zant}

\affiliation{Kavli Institute of Nanoscience Delft, Delft University of Technology,
Lorentzweg 1, 2628 CJ Delft, The Netherlands}

\date{19 December, 2005}

\begin{abstract}
We have fabricated longitudinal nanoconstrictions in the charge-density
wave conductor (CDW) NbSe$_{3}$ using a focused ion beam and using
a mechanically controlled break-junction technique. Conductance peaks
are observed below the T$_{P1}$$=145\,$K and T$_{P2}$$=59\,$K
CDW transitions, which correspond closely with previous values of
the full CDW gaps $2\Delta_{1}$ and $2\Delta_{2}$ obtained from
photo-emission. These results can be explained by assuming CDW-CDW
tunneling in the presence of an energy gap corrugation $\epsilon_{2}$
comparable to $\Delta_{2}$, which eliminates expected peak at $\Delta_{1}+\Delta_{2}$.
The nanometer length-scales our experiments imply indicate that an
alternative explanation based on tunneling through back-to-back CDW-normal
junctions is unlikely.
\end{abstract}



\maketitle
Charge-density wave (CDW) conduction remains of major interest despite
its experimental discovery nearly 30 years ago. Much of the existing
work has focused on transport properties of as-grown single crystals
\cite{gruener}. More recently, micro/nanofabrication methods for
CDW materials has allowed the study of mesoscopic CDW physics \cite{sinchenko,latyshev,zaitsev}.
Structures for tunneling spectroscopy are of particular interest because
of the unusual gap structure with large one-dimensional fluctuation
effects expected in these highly anisotropic materials, and because
of predictions of unusual mid-gap excitations of the collective mode
\cite{brazovskii}. Tunneling studies in fully gapped CDW conductors
like the \char`\"{}blue bronze\char`\"{} K$_{0.3}$MoO$_{3}$ suffer
from band-bending effects at the interface akin to semiconductor-insulator-metal
junctions. These effects are absent in the partially gapped CDW conductor
NbSe$_{3}$, which remains metallic down to $4.2\,$K.

Tunneling perpendicular to the direction $b$ of quasi-one-dimensional
chains, along which the CDW wavevector lies, has been studied in ribbon-like
whiskers of NbSe$_{3}$ by Scanning Tunneling Microscopy (STM) \cite{dai},
by lead contacts evaporated over the native oxide on the $b-c$ plane
\cite{fournel,sorbier}, and by tunneling through a gold wire or a
NbSe$_{3}$ crystal that is laid across another NbSe$_{3}$ crystal,
forming junctions in the $a-b$ or $b-c$ planes \cite{ekino87,ekino94}.
Peaks in the T=$4.2\,$K differential conductance at $35$ and $101\,$mV
\cite{dai}, $35\,$mV \cite{fournel}, $36\,$mV and $90\,$mV \cite{ekino87},
and $37\,$mV and $100\,$mV \cite{ekino94} from metal-NbSe$_{3}$
junctions correspond well with the CDW gaps $\Delta_{1}=110\,$mV
and $\Delta_{2}=45\,$mV for NbSe$_{3}$'s $T_{P1}=145\,$K and $T_{P2}=59\,$K
CDWs as determined by angle-resolved photo-emission (ARPES) \cite{schaefer}.
Crossed NbSe$_{3}$-NbSe$_{3}$ crystals \cite{ekino87} yielded peak
voltages of $60\,$mV and $142\,$mV, and interlayer tunneling in
micro-fabricated NbSe$_{3}$ mesas yielded peaks at $50\,$mV and
$120\,$mV \cite{latyshev}. A single in-chain tunneling study \cite{sinchenko}
using a gold ribbon mechanically positioned near the end of a NbSe$_{3}$
crystal gave a peak at $100\,$mV for the $T_{P1}$ CDW.

\begin{figure}
\begin{center}\includegraphics[%
  clip]{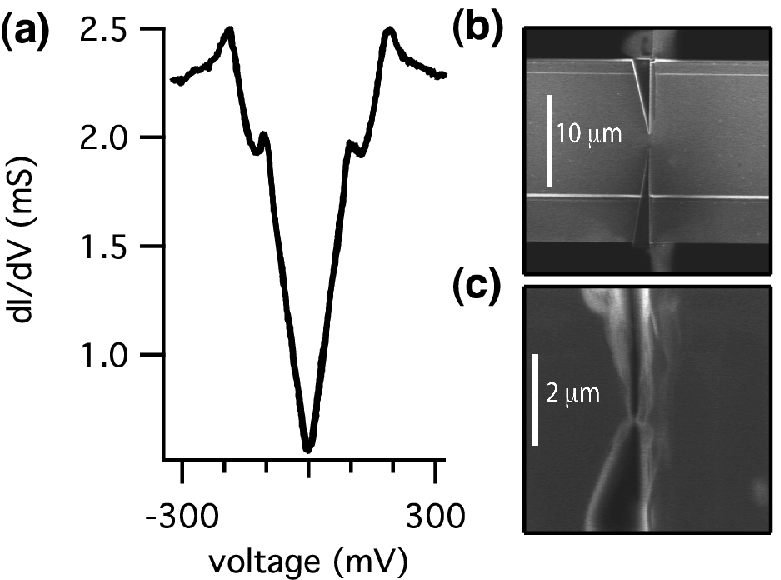}\end{center}

\caption{(a) Differential conductance $\frac{dI}{dV}$ vs applied voltage
at T$=4.2\,$K of a FIB-fabricated NbSe$_{3}$ nanoconstriction. (b)
and (c) show images taken in the FIB during fabrication.}

\label{graph:fig_1}
\end{figure}

Here we demonstrate that a small constriction in a NbSe$_{3}$ single
crystal, produced by dry etching with a Ga Focused Ion Beam (FIB),
shows conductance peaks at $106\,$mV and $190\,$mV corresponding
to $2\Delta_{1}$ and $2\Delta_{2}$, as illustrated in Figure \ref{graph:fig_1}.
We reproduce the data at $4.2\,$K using a Mechanically Controlled
Break-Junction (MCBJ) technique, demonstrating that the FIB results
are not dominated by Ga ion damage. Our results can be explained by
CDW-CDW tunneling in the presence of a large transverse gap corrugation,
although tunneling through back-to-back CDW-normal junctions cannot
be conclusively ruled out.

CDWs form in metals with quasi-one-dimensional Fermi surfaces. Electron-hole
pairs near the Fermi level $k_{F}$ form a macroscopic condensate
and associated periodic modulations of the electron density and atomic
positions. The condensate arises from the electron-phonon interaction,
as described by the mean-field Hamiltonian $H_{P}=\sum_{k\sigma}\epsilon_{k}a_{k\sigma}^{\dagger}a_{k\sigma}-\sum_{k\sigma}\left(a_{k\sigma}^{\dagger}a_{k+k_{F}\sigma}\Delta e^{i\phi}+h.c.\right)$
\cite{froehlich_peierls}, where $a_{k\sigma}$ $\left(a_{k\sigma}^{\dagger}\right)$
is the electron creation (annihilation) operator for the states with
wavevector $k$ and spin $\sigma$. Like conventional superconductors,
the CDW condensate produces peaks in the density-of-states (DOS) at
$\pm\Delta$ relative to the Fermi energy. Applying the semiconductor
model for electron tunneling in superconductor junctions \cite{blonder}
to CDW systems, tunneling between occupied and unoccupied states in
a CDW-insulating-CDW junction should produce peaks in the conductance
at voltages equal to $\pm\frac{2\Delta}{e}$. In a CDW-normal junction,
peaks should be observed at voltages equal to $\pm\frac{\Delta}{e}$.

To fabricate nanoconstrictions using the FIB, a single-crystal whisker
of NbSe$_{3}$ with a typical width of $20\,\mu$m was placed on a
silicon dioxide/silicon substrate having $2\,\mu$m wide Au electrical
contacts that were pre-patterned by photo-lithographic techniques.
The NbSe$_{3}$ crystal was then carved using an FEI/Philips FIB-200
focused ion beam. At low magnification ( $10,000\times$), two large
transverse cuts were made from either side to create a constriction
in the $\mathbf{b}/\mathbf{b^{\star}}$ direction. At high magnification
($50,000\times$), further line cuts were made at low currents of
$350\,$pA until the constriction had a width of around $100\,$nm
and its resistance (which dominated the overall sample resistance)
exceeded $150\,\Omega$ at room temperature. Images of both cuts,
taken in the FIB, are shown in Figure \ref{graph:fig_1}(b). To find
the cross-sectional dimension of the junction we can make a simple
estimate using the classical (diffusive) resistance formula, the contact's
room temperature resistance of $165\,\Omega$, NbSe$_{3}$'s bulk
resistivity ($1.86\,\Omega\mu$m), and a contact length of around
$100\,$nm (estimated from the device image in Figure \ref{graph:fig_1}(b)),
which gives a conducting area of roughly $20 $nm$^{2}$.

\begin{figure}
\begin{center}\includegraphics[%
  clip]{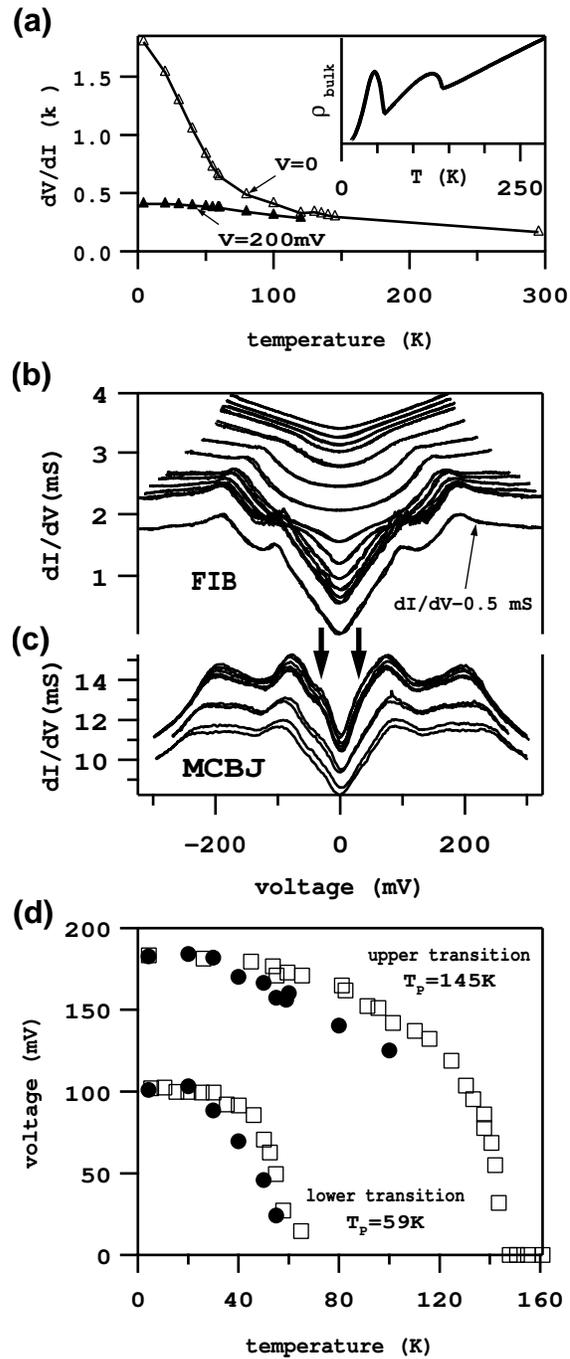}\end{center}

\caption{(a) $\frac{dI}{dV}$ of the FIB nanoconstriction at $V=0\,$mV and
$V=200\,$mV as a function of temperature. Anomalies due to the two
Peierls transitions at $145\,$K and $59\,$K seen in the bulk resistivity
(inset) are absent. (b) $\frac{dI}{dV}$ vs voltage of the FIB constriction
at temperatures (top to bottom) of $4.2\,$K, $20\,$K, $30\,$K,
$40\,$K, $50\,$K, $80\,$K, $100\,$K, $120\,$K, $130\,$K, $135\,$K,
$140\,$K and $145\,$K. The $4.2\,$K data is repeated, offset downward
by $-0.5\,$mS, for clarity. (c) $\frac{dI}{dV}$ vs voltage of the
break-junction sample taken at different stresses/separations. (d)
Temperature dependence of peak positions obtained from the FIB constriction
data of (b) (filled circles) compared with X-ray diffraction experiments
of the CDW order parameter \cite{fleming} (open squares).}

\label{graph:fig_2}
\end{figure}

Differential conductance versus voltage data was measured between
T$_{P1}=145\,$K and $4.2\,$K using a conventional four-probe technique
\cite{footnote_samples}. Figure \ref{graph:fig_2}(a) shows that
the constriction's zero-bias resistance increases monotonically with
decreasing temperature, and at $4.2\,$K has a value close to $2\,$k$\Omega$.
This contrasts with the bulk behavior of NbSe$_{3}$, which shows
large resistance increases just below the two Peierls transitions
and then a strong metallic (roughly linear) decrease down to $4.2\,$K
(inset).

To fabricate constrictions by the mechanically-controlled break-junction
technique, a single crystal whisker of NbSe$_{3}$ was placed on a
Kapton-tape capped piece of flexible phosphor-bronze. The crystal
was held to the Kapton, which had pre-patterned gold gold electrical
contacts, using cellulose. The crystal was controllably broken at
$4.2\,$K in a custom-built cryostat \cite{ruitenbeek}, which allowed
the sample to be broken and re-contacted several times in the course
of an experiment. Transport measurements at $T=4.2\,$K were performed
in two-probe configuration, with a contact resistance of $\sim10\,\Omega$
estimated from the total sample plus contact resistance before breaking.
Because of the quasi-one-dimensional bonding and very strong bonds
along the chains, the NbSe$_{3}$ crystal's response to stress likely
involved successive breaking of fibers within its cross-section, as
in the breaking of a rope. 

Figure \ref{graph:fig_2}(b) shows the differential conductance of
the FIB sample as a function of voltage for several temperatures,
with the lowest temperature differential conductance offset downwards
for clarity. At $T=4.2\,$K, peaks occur at $\pm105\,$meV and $\pm190\,$meV,
symmetric around zero bias, with a width $\sim30\,$meV and an estimated
error $\sim6\,$meV. The peak positions move to smaller voltage with
increasing temperature. Above its corresponding $\sim2T_{P}/3$, each
peak becomes an inflection, and above $T_{P}$ each inflection disappears,
indicating a strict association of each peak with each Peierls instability.
Figure \ref{graph:fig_2}(c) shows the corresponding data at $T=4.2\,$K
from the MCBJ sample. Peaks are clearly visible at $\pm81\,$mV and
$\pm196\,$mV.

The agreement between the MCBJ and FIB samples indicate that the transport
properties of the FIB samples are not qualitatively changed by any
damage or disorder caused by Gallium atoms. The measured $T=4.2\,$K
peak positions correspond well to recent angle-resolved photo-emission
spectroscopy measurements of the full CDW gaps $2\Delta_{1}$ and
$2\Delta_{2}$ of $90$ and $220\,$meV, respectively, \cite{schaefer}.
Figure \ref{graph:fig_2}(d) shows the position of the conduction
peaks (or inflections) with temperature. Their temperature dependence
closely matches that of the CDW order parameter determined from X-ray
diffraction measurements \cite{fleming}. Small sub-gap features are
also observed in both the FIB and more clearly in the MCBJ samples
(indicated by arrows), possibly due to soliton effects \cite{brazovskii}.
Careful analysis of Fig. \ref{graph:fig_2}(b) shows the absence of
any feature at $\frac{\Delta_{1}+\Delta_{2}}{e}\simeq145\,$meV, which
might be expected for tunneling between the $T_{P1}$ CDW on one side
of the constriction and the $T_{P2}$ CDW on the other. 

\begin{figure}
\begin{center}\includegraphics[%
  clip]{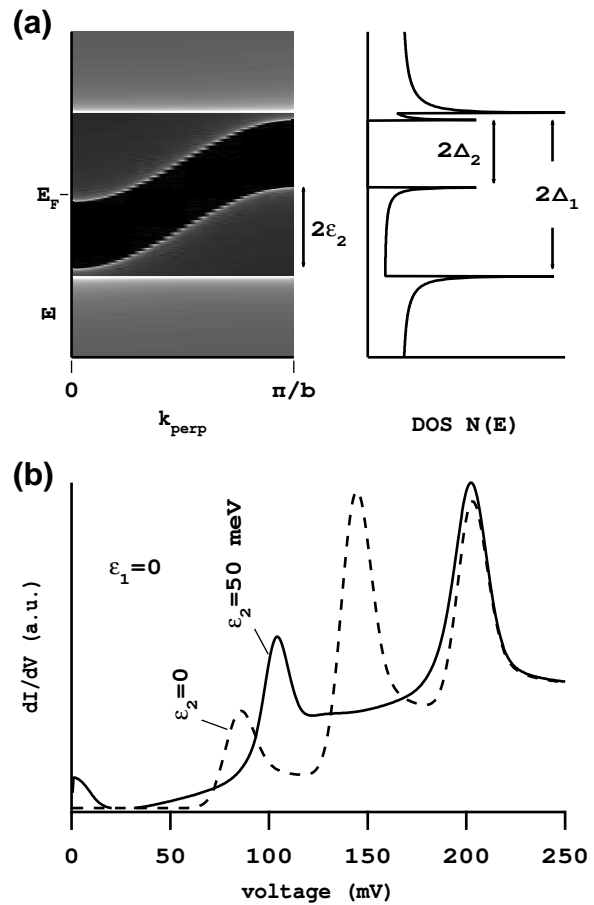}\end{center}

\caption{Semiconductor model for transitions between quasi-particle excited
states of the CDW, with corrugation $\epsilon_{2}$ present in the
transverse momentum direction of the $T_{P2}=59\,$K but not the $T_{P1}=145\,$K
CDW states. (a) \emph{Left}: Density of states $N(E,k_{\perp})$ as
a function of perpendicular wavevector and energy, showing how corrugation
of the $T_{P2}$ CDW changes the relative positions of the DOS maxima.
Brighter regions denote higher DOS. \emph{Right}: Projection of $N(E,k_{\perp})$
at a fixed wavevector $k_{\perp}=\frac{\pi}{b}$. (b) Simulated $\frac{dI}{dV}$
vs. $V$ using a BCS density of states broadened by $24\,$meV, and
corrugations $\epsilon_{2}=0$ and $50\,$meV. All calculations assume
$\Delta_{1}=100\,$meV and $\Delta_{2}=41\,$meV.}

\label{graph:fig_3}
\end{figure}

The conductance peak widths do not significantly vary below $\frac{T_{P}}{2}$
for either transition, implying a temperature-independent intrinsic
broadening of the CDW DOS. Down to the lowest temperatures, the conductance
also shows a substantial tail as the voltage decreases below the gap,
indicating that the DOS does not cut off sharply at the nominal gap
energy. This is consistent with optical absorption measurements \cite{perucchi},
and calculations of the zero-temperature CDW DOS with one-dimensional
fluctuations \cite{kim_mckenzie}. Suppression of conductance peaks
at $>2T_{P}/3$ for each CDW is likely due to the strong effect of
thermal fluctuations on the DOS in a quasi-one-dimensional system
\cite{lee}. 

Our results may be explained in terms of CDW-CDW tunneling in the
presence of a transverse corrugation of the $T_{P2}$ CDW's energy
gap, as shown in Figure \ref{graph:fig_3}. Inter-chain coupling results
in band structure dispersion in momenta perpendicular to the chains
($b-$direction), which should produce transverse corrugations in
the CDW gap and imperfect nesting \cite{huang}. In NbSe$_{3}$, the
measured electrical anisotropy implies a bandwidth in the $c-$direction
a factor of $10$ larger than in the $a-$direction, qualitatively
consistent with band structure calculations \cite{schaeferpriv},
so that the former should dominate in tunneling involving the $T_{P2}$
gap. Analysis of measurements on metal-NbSe$_{3}$ tunnel junctions
at low biases \cite{sorbier} and of bulk low-temperature thermal
and electrical conduction \cite{mihaly} have both suggested a transverse
bandwidth comparable to the $T_{P2}$ gap itself. 

We consider a semiconductor tunneling model \cite{blonder} including
finite transverse dispersion in the $c-$direction. This dispersion
is characterized by a single parameter $\epsilon_{0}=\frac{t_{\bot}^{2}cos\left(bk_{F}\right)}{2t_{b}sin\left(bk_{F}\right)}$,
$t_{\bot}$ and $t_{b}$ being the bandwidths of the dispersion perpendicular
and parallel to the chains, and $b$ the inter-chain separation. Figure
\ref{graph:fig_3}(a) illustrates this dispersion, showing by brightness
the density of states $N(E,k_{\perp})$ as a function of energy and
perpendicular wave vector within one half Brillouin zone (left), and
the density of states at a $N(E)$ at a fixed wavevector (right).
The conductance is then calculated by integrating the product $N(E,k_{\perp})N(E-eV,k_{\perp})$
over all energies $E$ and wavevectors $k_{\perp}$, assuming a BCS
$T=0$ DOS broadened by a Gaussian distribution of width $\sigma=24\,$meV
to mimic the experimental peak broadening. The computed $\frac{dI}{dV}(V)$
curves for $\epsilon_{2}=0$ (no corrugation) and $\epsilon_{2}=50\,$meV
are shown in Figure \ref{graph:fig_3}(b). As expected, with no corrugation
$\frac{dI}{dV}(V)$ exhibits a strong peak at $(\Delta_{1}+\Delta_{2})/e$.
As $\epsilon_{2}$ is increased, this peak splits in energy and shrinks
in height. While all conductance peaks remain visible with a divergent
DOS, with a broadened DOS the peaks at $\Delta_{1}+\Delta_{2}-\epsilon_{2}$
and $2\Delta_{2}$ merge at around $100\,$meV, resulting in a conductance
curve that closely resembles the present data. The transverse gap
corrugation we assume to obtain the best fit to our data is in agreement
with the conclusion of Sorbier \emph{et al.} \cite{sorbier} that
the corrugation for the $T_{P2}$ CDW $\epsilon_{2}$ is slightly
larger than the CDW gap $\Delta_{2}$.

An alternative explanation is that the peak structure results from
back-to-back N-CDW junctions. Provided that the relaxation length
of the non-equilibrium distributions is long compared with the characteristic
length of the transition from CDW to normal states, one would expect
$\frac{dI}{dV}$ peaks corresponding to $2\Delta_{1}$ and $2\Delta_{2}$,
and no intermediate peak. We can estimate the dimensions of the constricted
region assuming that the narrow region consists of NbSe$_{3}$ chains.
The length of the constriction can be determined by comparing the
low-field residual resistance ratio $\frac{R(T=293 K)}{R(T=4.2 K)}$
with NbSe$_{3}$ nanowires of different cross-sectional dimensions
\cite{slot}. We find that our FIB junction most closely resembles
a wire with resistance per unit length $R/L\sim10^{6} \Omega/\mu$m
at $4.2 $K, corresponding to a cross-section of $600\,$nm$^{2}$.
From the measured junction resistance $R(T=4.2 K)=1798 \Omega$ we
obtain a junction length $\sim1.8\,$nm. This is comparable to the
known CDW amplitude coherence length in NbSe$_{3}$, and is too short
to allow the full loss of CDW order required to achieve the tunneling
characteristics of back-to-back CDW-normal junctions. Furthermore,
despite the strong zero-bias low-temperature resistance increase,
the FIB constriction shows no hints of anomalies at the two Peierls
transitions. These are clearly seen even in the smallest cross-section
nanowires ($500 $nm$^{2}$) studied to date. 

The back-to-back junction interpretation might be viable if the FIB
somehow disorders the constricted region in a way that eliminates
the Peierls transitions and causes the resistance to increase strongly
with temperature, while producing a resistivity whose magnitude is
smaller than in nanowires having similar temperature dependence. But
even in this case the observed zero-bias junction resistance cannot
easily be accounted for \cite{footnote_b2b}. In addition, the relevance
of the back-to-back junction explanation to MCBJ samples, where the
sample has simply been broken and brought back close together is unclear
due to the absence of an intermediate conducting structure. We therefore
attribute the conductance peaks in our devices to CDW-CDW tunneling. 

In conclusion, we have fabricated in-chain nanoconstrictions in a
CDW material. These constrictions behave like a tunnel junctions,
and conductance peaks are observed at biases that correspond well
with the full CDW gaps $2\Delta_{1}$ and $2\Delta_{2}$ determined
from independent measurements. The peaks disappear at around two-thirds
the respective Peierls transition temperature, in agreement with calculations
of the effects of fluctuations on the DOS for one-dimensional compounds.

We thank S. Za\u{i}tzev-Zotov and S. Artemenko for fruitful discussions.
This work was supported by the International Association for the Promotion
of Co-operation with Scientists from the New Independent States of
the Former Soviet Union (INTAS-NIS), the Foundation for Fundamental
Research on Matter (FOM), and the National Science Foundation (NSF)
(Grants No. DMR 0101574 and No. INT 9812326). K.O'N. was supported
by the Marie Curie Fellowship organization. We thank J. van Ruitenbeek
for the use of equipment in the MCBJ work, and S. Otte and R. Thijssen
for technical assistance.


\begin{thebibliography}{20}
\bibitem{footnote_samples}Two other FIB samples with insulating properties showed features near
$2\Delta_{1}$ and $2\Delta_{2}$. One of these had inflection points
near $200 $mV and $106 $mV, the other had a peak at $170 $mV and
an inflection at $\sim100 $mV.
\end{thebibliography}
\end{document}